\documentstyle[12pt]{article}
\begin{document}
\input epsf.tex
\def\a{\alpha}
\def\b{\beta}
\def\ch{\chi}
\def\d{\delta}
\def\e{\epsilon}
\def\f{\phi}
\def\g{\gamma}
\def\h{\eta}
\def\i{\iota}
\def\j{\psi}
\def\k{\kappa}
\def\l{\lambda}
\def\m{\mu}
\def\n{\nu}
\def\o{\omega}
\def\p{\pi}
\def\q{\theta}
\def\r{\rho}
\def\s{\sigma}
\def\t{\tau}
\def\u{\upsilon}
\def\x{\xi}
\def\z{\zeta}
\def\D{\Delta}
\def\F{\Phi}
\def\G{\Gamma}
\def\J{\Psi}
\def\L{\Lambda}
\def\O{\Omega}
\def\P{\Pi}
\def\S{\Sigma}
\def\U{\Upsilon}
\def\T{\Theta}

\def\Ab{\bar{A}}
\def\gi{g^{-1}}
\def\li{{ 1 \over \l } }
\def\lb{\l^{*}}
\def\zb{\bar{z}}
\def\ub{u^{*}}
\def\Tb{\bar{T}}
\def\pp {\partial }
\def\pb {\bar{\partial }}
\def\be{\begin{equation}}
\def\ee{\end{equation}}

\def\phib{\phi^\dagger}
\def\beq{\begin{eqnarray}}
\def\eeq{\end{eqnarray}}
\def\ben{\begin{eqnarray}}
\def\een{\end{eqnarray}}
\def\Tr{{\rm Tr}}
\def\X{{\bf X}}
\def\sech{{\rm sech}}

\hsize=16.5truecm
\addtolength{\topmargin}{-0.6in}
\addtolength{\textheight}{1.5in}
\vsize=26truecm
\hoffset=-.6in

\baselineskip=7 mm
\rightline{\today} 
\vskip2cm
\centerline{\Large\bf Three-dimensional Curve Motions}
\centerline{\Large\bf Induced by the Modified Korteweg-de Vries Equation}

\vskip 1cm

\centerline{H.J. Shin\footnote{e-mail:
hjshin@khu.ac.kr}}

\vskip3mm
\centerline{Department of Physics and Research Institute of Basic
Science}
\centerline{Kyung Hee University, Seoul 130-701,  Korea}
\vskip 2cm
\centerline{\bf ABSTRACT}
\vskip 5mm

We have constructed one-phase quasi-periodic solutions of the curve equation induced by the mKdV equation. 
The solution is expressed in terms of the elliptic functions of Weierstrass. 
This solution can describe curve dynamics such as a vortex filament with axial velocity 
embedded in an incompressible inviscid fluid. There exist two types of curves (type-A, type-B) 
according to the form of the main spectra of the finite-band integrated solution. 
Our solution includes various filament shapes such as the Kelvin-type wave, 
the rigid vortex, plane curves, closed curves, and the Hasimoto one-solitonic filament.

\vskip 5mm

\newpage

\setcounter{footnote}{0}

\vskip 5mm

\section{Introduction}
Space curves are of interest in mathematics and physics. This is especially true with 
how these curves' mathematical structures are related to integrable equations. 
The first appearance of such a relationship was between the sine-Gordon equation 
and the differential geometry for surfaces of constant negative curvature \cite{embed}. 
Another interesting development was from Hasimito who showed the motion of 
vortex filament motion is related to the nonlinear Schr\"{o}dinger equation \cite{hasimoto}. 
Motivated by this development, Lamb studied various helical curves that were 
described by well-known integrable equations \cite{lamb}. He showed that there 
exists a curve that has a constant torsion and its curvature is described by the 
modified Korteweg-de Vries (mKdV) equation. 

A special class zero torsion curves are the planar curves, which already appeared in the famous 
old problem of elastica. The elastica problem was solved by Euler. Typical solutions 
of the elastica problem, found by a numerical method, can be seen in the book \cite{love}, 
which are essentially those solutions of Euler. After that, this problem has occurred 
repeatedly in diverse forms. For example, Ishimori \cite{ish} found that the soliton 
solution of the mKdV equation is essentially the same as the loop soliton solution 
found  in  \cite{kon}. The loop soliton described the loop solitary wave propagating 
along a stretched rope. Mumford showed that computer vision problems were related 
to the mKdV equation \cite{mumf}. A new form of the solution was given by Mumford  
which was expressed by the theta functions of the genus-one Riemann surface \cite{mumf}. 
More general solutions, corresponding to the genus-N Riemann surfaces, were given 
by Matsutani in terms of the Weierstrass sigma functions and their generalizations \cite{mats}. 
On the other hand, \cite{gold} and \cite{wadati} related the modified Korteweg-de Vries (mKdV) 
equation to motions of curves in a plane.

The above development of the elastica problem was confined to space curves having 
zero torsion and laying on a plane. On the other hand, 
the problem posed by Lamb, finding space curves related to the mKdV equation 
having a constant torsion and thus moving in three-dimensional space, was not 
investigated thoroughly in the research literature. As we can see in Section 2.4, 
this problem is related to the motion of a vortex filament with axial velocity and 
moving in an incompressible inviscid fluid \cite{fuku}. Generally, constructing a curve 
having a given curvature and torsion is a difficult problem. It can be carried out for 
a simple from of curvature and torsion only \cite{eise}. For example, curves having 
a curvature of the one-soliton and constant non-zero torsion are explicitly 
constructed in the Appendix of \cite{lamb}. Curves, having a more complex form 
of curvature, have been constructed for special cases only, by applying methods 
specially adopted to the given problem \cite{shin3}.

In our work, described in this paper, we construct space curves whose curvature 
is described by the quasi-periodic solution of the mKdV equation and curves 
which have constant non-zero torsion. The constructed curve is of the Kida-type 
stationary solution \cite{kida}. Equations that describe this Kida-type stationary 
solution were already known in \cite{fuku}, but were not solved because of their 
complicated form. The present calculation applied the method of the  
``modified  finite-band integration". The modified method was invented to solve 
the so-called effectivization problem, which is related to the extraction of the real solutions from the finite-band integration 
method \cite{tracy,kam}. This modified method gives solutions in the simplest but important 
one-phase case.
The modified method gave physical solutions in a rather 
simple form , which made this method useful to apply to real situations. 
Recently it was used to obtain quasi-periodic solutions of various integrable theories  
expressed by the Weierstrass elliptic functions \cite{kam1,shin4}. It was then applied to 
space curve problems, such as the Heisenberg model, filament motions under 
the localized induction approximation, and the Lund-Regge vortex motions 
\cite{kam2,shin1,shin2}. These problems were related to the nonlinear Schr\"{o}dinger 
equation and complex sine-Gordon equation, respectively. 

A recent related calculation 
was the  application of the Sym-Tafel formula to the problem of filament motions 
under the localized induction approximation \cite{cali}. Ref. \cite{cali} also obtained 
real quasi-periodic solutions  expressed in terms of the Riemann theta function. 
This approach, based on the standard finite-band integration method, has the merit of obtaining
more general solutions of multi-phase case. But these solutions through the higher-genus Riemann surfaces
have complicated form, including numerical integrations to explictly enumerate
parmeters appearing in the solutions.
Thus it was difficult for physicists to apply the quasi-periodic solutions of these type to
the analyses of experimental results. This was the reason of appearing one-phase solutions
of various integrable equations in the previous paragraph 
by using the modified finite-band integration methods. Note that there generally exist multi-phase
solutions of various integrable equations in terms of the Riemann theta function, though they
are not widely used in the analyses of physical problems \cite{kam1}. In fact, the present paper gives solutions
expressed in terms of the Weierstrass' elliptic fuctions with explicit parameters. 
Using the known properties of the Weierstrass' functions,
we can trace the variation of physical properties, like moving or rotation velocities, 
of the solutions along with the spectral parameters.

The derivation of the curve configurations, described in the present paper, is similar to 
the previous cases of \cite{kam2,shin1,shin2}. But there exists an important difference 
with the present calculation as compared to the previous ones, related with imposing 
conditions such as the constancy of torsion $\t$. It then results in a constraint on the 
main spectra  and classifies the calculated curves into two categories (type-A and type-B). 
This restriction on the main spectra is not observed in the previous calculations of the 
modified finite-band integration method \cite{kam1,shin1,shin2}. Solutions of a more simple type, 
discussed in \cite{fuku}, including the one-solitonic curve, the helicoidal vortex filament, 
and the Euler's elastica, are obtained by the reduction from the curve of the present 
calculation. 

The general setup for the curve problem of the mKdV equation is given in Section 2. 
Explicit construction of the one-phase periodic solution is followed using a modified version 
of the finite-band integration method in Section 3. Especially, imposing special conditions 
of the mKdV problem is given in Section 3.4. The resulting formulae involve Weierstrass' 
elliptic functions.
In Section 4, various specific solutions from the reduction of periodic solutions are explained 
using the functional relations of Weierstrass' functions. Some numerical plots, as well 
as an explicit check of solutions, are accomplished with the help of the symbolic package, 
Mathematica. Section 5 follows with a discussion.

\section{mKdV-type space curves}

\subsection{Equations of curve motion}

The space curve in three dimension is described by $\{\bf t, n, b\}$ vector (the unit tangent, 
normal, and binormal vectors, respectively). Here, ${\bf t}= \pp {\bf r} / \pp z$, 
where ${\bf r}(z, \zb)$ is the position vector of a point on the curve parametrized by the arc length $z$
at time $\zb$. They satisfy the following Serret-Frenet equation,
\be
\pp {\bf t}= \k {\bf n},~\pp {\bf n}= -\k {\bf t} +\t {\bf b},~ \pp {\bf b} = -\t {\bf n},
\label{FS}
\ee
where $\pp \equiv {\pp /  \pp z}$ and $\k, \t$ are respectively the curvature and torsion.
Generally, motion of a point on the curve can be specified in the form
\be
\dot {\bf r} \equiv \pb {\bf r} = U {\bf n} + V {\bf b} +W {\bf t},
\label{dotr}
\ee
where $\pb \equiv {\pp  / \pp \zb}$ is the time derivative.
Especially, the mKdV-type curves with a constant torsion $\t$ is described by \cite{lamb}
\be
U=-\pp \k, ~V=2 \k \t,~W=3 \t^2 -\k^2/2.
\label{UVW}
\ee

The family of curves with $\t=0$ are shown to be related with the problem of arc length conserving (nonstretching) 
motion and/or area conserving motion on a plane \cite{gold,wadati}. It was first analyzed by Euler in his study 
of the thin rod, the so-called the elastica problem. Various possible shape of curves are illustrated
in \cite{love}, including the famous figure-8 shape. 
These type of solutions are 
obtained by assuming that curves lye on a plane, thus leading to curves of zero torsion.
On the contrary, the interesting problem of non-zero torsion has not been dealt in depth
in literature and only the simple solutions has been obtained \cite{lamb}.

\subsection{Integrable structure of the mKdV equation}
We first start with the  zero curvature condition of the mKdV equation,
\be
[~ \pp +U_{KV} , \ \pb  +V_{KV} ~ ] = 0 ,
\label{zero}
\ee
where 
\ben
U_{KV} &=&  \pmatrix{0 & \k/2 \cr -\k/2 & 0 } + i \t /2 \pmatrix{1 & 0 \cr 0 & -1}, \nonumber \\
V_{KV} &=& \pmatrix{0 & -(\pp^2 \k ) /2 -\k^3 /4 \cr 
(\pp^2 \k )/2 +\k^3/4 & 0 } +i \t \pmatrix{- \k^2 /4 &  (\pp \k )/2 \cr   (\pp \k )/2 &  \k^2 /4 } \nonumber \\
&+& \t^2
\pmatrix{ 0 & \k/2 \cr -\k/2 & 0} +i \t^3 /2 \pmatrix{1 & 0 \cr 0 & -1}.
\een
The mKdV variable $\k=\k(z, \zb)$ becomes the curvature of the corresponding curve problem,
while the spectral parameter $\t$ becomes the torsion.
Note that $\t$ is a constant and it does not depend on $z, \zb$.
The zero curvature condition in Eq. (\ref{zero}) is the compatibility condition for 
the overdetermined system of the following associated linear equations
\be
(\pp + U_{KV} )\Phi = 0 ,\ \
(\pb + V_{KV} )\Phi = 0.
\label{linear}
\ee
By explicitly enumerating the zero curvature condition at the level of $O(\t^0)$, we obtain the mKdV equation,
\be
\pb \k+\pp^3 \k +{3 \over 2} \k^2 \pp \k=0,
\label{mkdv}
\ee
while other levels of $O(\t^i), i=1,4$ just becomes identities.

\subsection{The Sym-Tafel formula}
Now we introduce the duality relation between the curve problem and the mKdV system.
We write down the 
curve position vector ${\bf r} \equiv (r_1, r_2,r_3 )$  in terms
of $\Phi$ in Eq. (\ref{linear}) and shows that the linear equation in  Eq. (\ref{linear})
implies the Serret-Frenet equation in Eq. (\ref{FS}). 
For this, we introduce the following Sym-Tafel formula,
\be
r  = 2i \Phi ^{-1} {\pp \over \pp \t} \Phi,
\label{fdef}
\ee
where $r \equiv \sum r_i \s_i,  i=1,3$,
with the Pauli matrices $\s_i$ \cite{sym,cies}. (From now on, we will omit the summation notation.)

To check that Eq. (\ref{fdef}) gives the Serret-Frenet equation, we first notice that
\be
\pp r = -2 i \Phi ^{-1} \pp \Phi \Phi ^{-1} {\pp \over \pp \t} \Phi + 2 i
\Phi ^{-1} {\pp \over \pp \t} (\pp \Phi) = -2i \Phi ^{-1} {\pp U_{KV} \over \pp \t} \Phi
= \Phi ^{-1} \s_3 \Phi.
\ee
In other words, the unit tangent vector ${\bf t} = {\pp {\bf r}} \equiv (t_1 , t_2 , t_3 )$ is given by 
\be
t = t_i \s_i = {\pp r_i } \s_i = {\pp r}= \Phi ^{-1} \s_3 \Phi.
\label{teqs}
\ee
Note that $t_i t_i = {1 \over 2} \Tr t^2 = 1.$
This equation shows that the tangent vector $\bf t$ is given by the rotation of
$\hat k$ (the unit vector along the $z$-axis), where the rotation is induced 
by the similarity transformation of $\Phi$.
In a similar way, we find that 
\be
{\pp t}=\pp t_i \s_i =\pp^2 r= -[ \Phi ^{-1} \pp \Phi,~\Phi ^{-1} \s_3 \Phi] = \Phi^{-1} [U_{KV},~\s_3] \Phi= -\k \Phi ^{-1} \s_1 \Phi.
\label{tn}
\ee
Eq. (\ref{tn}) is one of  the Serret-Frenet equation, $\pp \bf t =\k \bf n$ ($\k$ is the curvature), 
when we define the normal vector ${\bf n} \equiv ( n_1 , n_2 , n_3 )$ as
\be
n=n_i \s_i = - \Phi ^{-1} \s_1 \Phi.
\label{neqs}
\ee
Thus, the normal vector $\bf n$ is given by the rotation of 
$\hat i$ (the unit vector along the $x$-axis). It is clear that
$\bf t$ and $\bf n$ are orthogonal to each other.
Now,
\be
\pp n =\pp n_i \s_i = -\pp (\Phi ^{-1} \s_1 \Phi)=[ \Phi ^{-1} \pp \Phi,~\Phi ^{-1} \s_1 \Phi]
=-\k \Phi ^{-1} \s_3 \Phi +\t  \Phi ^{-1} \s_2 \Phi,
\ee
which gives another Serret-Frenet equation $\pp \bf n = \t \bf b -\k \bf t$,
when we take the binormal vector ${\bf b} \equiv (b_1 , b_2 , b_3 )$ as 
\be
b \equiv b_i \s_i =  \Phi ^{-1} \s_2 \Phi.
\label{beqs}
\ee
In this case, the binominal vector $\bf b$ is given by the rotation of $\hat j$ (the unit vector along the $y$-axis). 
Finally, similar calculation gives
\be
\pp b =\pp b_i \s_i = \pp (\Phi ^{-1} \s_2 \Phi)=-[ \Phi ^{-1} \pp \Phi,~\Phi ^{-1} \s_2 \Phi]
=\t \Phi ^{-1} \s_1 \Phi ,
\ee
which gives another Serret-Frenet equation, $\pp \bf b = - \t \bf n$.

The time dependence of the $r$ is similarly constructed as
\ben
\pb r&=&-2 i \Phi ^{-1} \pb \Phi \Phi ^{-1} {\pp \over \pp \t} \Phi + 2 i
\Phi ^{-1} {\pp \over \pp \t} (\pb \Phi) = -2i \Phi ^{-1} {\pp V_{KV} \over \pp \t} \Phi \nonumber \\
&=&\Phi ^{-1} (3 \t^2 \s_3+2 \k \t \s_2-{1 \over 2} \k ^2 \s_3+\pp \k \s_1 ) \Phi \nonumber \\
&=& 3 \t^2 {t} + 2\k \t {b} -{1 \over 2} \k^2 {t}-\pp \k {n},
\label{pbX}
\een
where $t$ and $b$ are given by Eqs. (\ref{teqs}), (\ref{neqs}) and (\ref{beqs}).
It is just the Eq. (\ref{UVW}) and is already found by Lamb in \cite{lamb}.
This equation can be rewritten in terms of the curve variable $\bf r$, 
\be
\pb {\bf r} 
=-\pp^3 {\bf r}+\left({3 \over 2} |\pp^2 {\bf r} |^2 - 3 \t^2 \right) \pp {\bf r} 
+3 \t \pp^2 {\bf r} \times \pp {\bf r}
.
\label{nlssp}
\ee

\subsection{Equation for localized induction approximation with axial velocity}
Eq. (\ref{nlssp}) is the curve equation induced from the mKdV equation with the
spectral parameter $\t$. This equation can be rewritten to the equation for localized 
induction approximation generalized to take account of the axial-flow effect.
Consider the following change of coordinates,
\be
z \rightarrow X-{c^2 \over 3} T, ~~\zb \rightarrow T, ~~~\t \rightarrow -{c \over 3}
\label{trans}
\ee
Above coordinates change transforms Eq. (\ref{nlssp}) into the following form,
\be
\pp_T {\bf r} = -\pp_X ^3 {\bf r} -{3 \over 2} |\pp_X {\bf r}|^2 \pp_X {\bf r}
 -c \pp_X ^2 {\bf r} \times \pp_X {\bf r},
\label{eqnr}
\ee
where $\pp_T \equiv {\pp \over \pp T}, ~~\pp_X \equiv {\pp \over \pp X}$.
Eq. (\ref{eqnr}) is the equation introduced in \cite{fuku} and it describes
the motion of a thin vortex filament with axial velocity, embedded in a inviscid incompressible fluid. 
In \cite{fuku}, solutions
of Eq. (\ref{eqnr}) were introduced including the N-soliton solution, a circular helix and a plane curve
of Euler's elastica.
But the Kida-type solution was not given explicitly and only equations describing the  
Kida-type curve was presented. Our results can be used to fill up this gap by using the 
transformation in Eq. (\ref{trans}).

\subsection{Integrable structure of the curve equation}

We now use a modified form of the R-transformation \cite{orf,shin1,shin2} to obtain
the Lax pair of the curve equation. It will be used to obtain a quasi-periodic solution by applying the finite 
integration method in the next section.
First we define $\Psi \equiv \Phi^{-1}  \hat \Phi$
where $\Phi$, $\hat \Phi$ are solutions of the linear equation (\ref{linear})
with the spectral parameter $\t$, $\hat \t$, respectively . Define
\ben
M &\equiv& \pp \Psi \Psi^{-1}= -\Phi^{-1} \pp \Phi +\Phi^{-1} \pp {\hat \Phi} {\hat \Phi}^{-1} \Phi 
= \Phi^{-1} U_{KV} (\t) \Phi  - \Phi^{-1} U_{KV} ({\hat \t }) \Phi
\nonumber \\
&=& {i \over 2}  (\t  - \hat \t) \Phi^{-1} \s_3  \Phi = {i \over 2} (\t - \hat \t ) \pp r,
\nonumber \\
N &\equiv& \pb \Psi \Psi^{-1}={i \over 2} (\t^3 - \hat \t^3 ) \Phi^{-1} \s_3 \Phi +{1 \over 2} (\t^2-\hat \t^2) \Phi^{-1} 
\pmatrix{0 & \k \cr -\k & 0} \Phi \nonumber \\
&+&{i \over 4} (\t-\hat \t) \Phi^{-1} \pmatrix{-\k^2 & 2 \pp \k \cr 2 \pp \k & \k^2} \Phi
\nonumber \\
&=&{i \over 2} (-2 \t^3 +3 \hat \t \t^2 - \hat \t^3 ) \pp r
+{1 \over 4} (\t^2 - 2 \hat \t \t+\hat \t^2) [\pp r, ~\pp^2 r]+{i \over 2} (\t-\hat \t ) \pb r.
\een
Note that $M \Psi = \pp \Psi, ~N \Psi = \pb \Psi$. 
Now, we introduce $\l\equiv\hat \t-\t$, which becomes a new spectral parameter of the following
linear equation for $r$,
\be
(\pp - M) \Psi =0,~~ (\pb -N) \Psi =0,
\label{newlinear}
\ee
where
\ben
M &=& -{i \over 2} \l \pp r, \nonumber \\
N &=& -{i \over 2} \l^3 \pp r -i{3 \over 2} \l^2 \t \pp r + {1 \over 4} \l^2 [\pp r,~ \pp^2 r]-{i \over 2} \l \pb r.
\label{MN}
\een

Note that the compatibility
condition of the associated linear equation, i.e., $[\pp -M,~\pb-N] =0$ gives the following equation of motion for $r$,
\be
i{3 \over 2} \t \pp^2 r -{1 \over 4} [\pp r,~ \pp^3 r] -{1 \over 4} [\pp r,~\pb r]=0,
 \label{vortex}
\ee
as well as the  identity $\Tr (\pp r)^2 =2$.
Eq. (\ref{vortex}) gives $\bf n$ and $\bf b$ components of $\pb r$ in Eq. (\ref{dotr}),
which are $U=-\pp \k, V=2 \k \t$ in Eq. (\ref{UVW}). But Eq. (\ref{vortex})
does not give the $\bf t$ components of $\pb r$, i.e., $W=3 \t^2-\k^2/2$  in Eq. (\ref{UVW}).
This fact is not surprising since
$W$ is known not to be related with the intrinsic form of the space curve, and is related with the internal parametrization of
the curve \cite{wadati}. In this respect, Eq. (\ref{vortex}) just describes the intrinsic structure of the curve.
In the next section, we will derive the quasi-periodic solution of the mKdV-type curve, starting
from the associated linear equation in Eq. (\ref{newlinear}). In the course of derivation, we will additionally
require the condition $W=3 \t^2-\k^2/2$, as well as the constancy of $\t$, see section 3.4.

\section{One-phase quasi-periodic solution}
\subsection{Squared wavefunctions}
First we introduce notations,
\be
\pp r = \pmatrix{\a_3 & \a_+ \cr \a_- & -\a_3}, \
\pb r = \pmatrix{\b_3 & \b_+ \cr \b_- & -\b_3}, \ [\pp r,~ \pp^2 r]=\pmatrix{\g_3 & \g_+ \cr \g_- & -\g_3},
\label{UV}
\ee
where $\a_+ ^* =\a_-,~ \a_3 ^* =\a_3, ~ \b_+ ^* =\b_-, ~\b_3 ^* = \b_3, \g_+ ^* =-\g_-,~ \g_3 ^* =-\g_3.$
The equation of motion in Eq. (\ref{vortex}) is rewritten as
\ben
6i \t \pp \a_3 -\pp \g_3 &=&  \a_+  \b_-  - \a_-  \b_+  ,\nonumber \\
3i \t \pp \a_+ -{1 \over 2} \pp \g_+ &=&  \a_3  \b_+  - \a_+  \b_3,
\een
while the constraint $\Tr (\pp r )^2 = 2$ becomes $\a_3 ^2 +\a_+ \a_-  =1$.

The finite-band integration method relies on the fact that the curve satisfying Eq. (\ref{vortex})
guarantees the existence of the solution $\Psi$ of Eq. (\ref{newlinear}) as a function of $\l$. 
The following procedure assumes that the squared function of $\Psi$ is a  polynomial function in $\l$, and thus
is called by the ``modified squared wavefunctions method" \cite{kam}.
Let the 
systems (\ref{newlinear}) have two basic solutions of column matrices, $\Psi_1 =(\j_1, \j_2)$ and
$\Psi_2 =(\f_1, \f_2)$, which is used to build a squared wave function such that
\be
f=-(i/2) (\j_1 \f_2 + \j_2 \f_1),~~g=\j_1 \f_1,~~h=-\j_2 \f_2.
\label{fgh}
\ee
From the definition of $f, g$ and $h$ in Eq. (\ref{fgh}) and using Eqs. (\ref{newlinear}), (\ref{MN}) and (\ref{UV}),
we can obtain following equations,
\ben
\pp f &=& -{\l \over 2} \a_- g  +{\l \over 2} \a_+ h,~~~~~\pp g = \l \a_+ f -i \l \a_3 g, \nonumber \\
\pb f &=& ({1 \over 2} \l^3 \a_+ +{3 \over 2} \l^2 \t \a_+ + {i \over 4} \l^2 \g_+ +{1 \over 2} \l \b_+ )h 
- ({1 \over 2} \l^3 \a_- +{3 \over 2} \l^2 \t \a_- +{ i \over 4} \l^2 \g_- +{1 \over 2} \l \b_- )g, \nonumber \\
\pb g &=& (-i \l^3 \a_3 -3i \l^2 \t \a_3 + {1 \over 2} \l^2 \g_3 -i\l \b_3 )g 
+ (\l^3 \a_+ +3 \l^2 \t \a_++{ i \over 2} \l^2 \g_+ +\l \b_+ )f.
\label{eqm}
\een
Equations for $h$ are similarly given.

Using Eqs. (\ref{eqm}), it can
be explicitly checked that $P(\l) \equiv f^2 -gh$ is independent of
$z$ and $\zb$ and is  only the function of $\l$. 
The so-called 1-phase periodic
solution is obtained by specially taking the form of 
\be
P(\l)= f^2 -gh = \prod_{i=1,4}  (\l-\l_i ),
\label{P}
\ee
where $\l_i$ are zeros of the polynomial which characterize the periodic 
solution of the ``Bloch wave" problem and are called the main spectra. 
The zeros $\l_i$ have to consist of complex conjugate pairs
$\l_j = \l_{Rj} +i \l_{Ij} , \l_{j+2} = \l_{Rj} -i \l_{Ij},~j=1,2$, which is required for that the obtained curve ${\bf r}$
is real. Eqs. (\ref{eqm}) and (\ref{P}) are
consistent with the following polynomial form of $f, g$ and $h$ in $\l$,
\be
f=\a_3 \l^2+ f_1 \l +f_0,~~g=-i\l \a_+  (\l-\m),
~~h=-i \l \a_- (\l-\m ^*),
\label{fgh1}
\ee
where $\m$ are functions of $z, \zb$.

\subsection{Equations for $\m,f_i, \a_i, \b_i, \g_i$}
Using Eqs. (\ref{fgh1}) and (\ref{eqm}), we can evaluate $\pp g$ at $\l=\m$ to find
\be
\pp \m = -i (\a_3 \m^2 +f_1 \m + f_0 ) .
\label{eq1}
\ee
If we look at $O(\l^{i}), i=2,1,0$ terms for $\pp f$ in Eq. (\ref{eqm}), we find
\be
\pp \a_3 =-{ i \over 2} \a_+ \a_-  (\m - \m ^*),~~~~~\pp f_1 =\pp f_0 =0 .
\label{eq11}
\ee
At $O (\l^{2} )$ terms for $\pp g$, we find
\be
\pp \a_+ = i \a_3 \a_+ \m  +i \a_+ f_1 .
\label{eqp}
\ee
Applying the same procedure to $\pb g$  and $\pb f$ in Eq. (\ref{eqm}), we can obtain that
\ben
i \m \a_+ \pb \m &=& (\m^3 \a_+ +3 \t \m^2 \a_+ +{i \over 2} \m^2 \g_+ +\m \b_+ )(\a_3 \m^2  +f_1 \m + f_0)
\nonumber \\
\pb \a_+&=&i \b_3 \a_+ \m-{1 \over 2} \g_+ f_0 +i \b_+ f_1 +3 i \t \a_+ f_0, \nonumber \\
0 &=& \a_3 \a_+ \m + \a_+ f_1 +{i \over 2} (\g_+ \a_3 -\g_3 \a_+ ), \nonumber \\
0 &=&{i \over 2} \g_3 \a_+ \m -\b_3 \a_+ +3 \t \a_3 \a_+ \m +{i\over 2} \g_+ f_1 + \a_+ f_0 +\b_+ \a_3 +3 \t \a_+ f_1,
\label{eqNN}
\een
and
\ben
\pb \a_3 &=& {i \over 2} (\b_+ \a_- \m^* -\b_- \a_+ \m), ~~~~~~\pb f_1 =\pb f_0 =0, \nonumber \\
0 &=&2 \a_+ \a_- (\m^* -\m) -i (\g_+ \a_- - \g_- \a_+ ), \nonumber \\
0 &=& i (\g_+ \a_- \m^* -\g_- \a_+ \m) -2 (\b_+ \a_- -\b_- \a_+ ) +6 \t \a_+ \a_- ( \m^* -\m).
\label{eqN}
\een

\subsection{$\m, f_i, \g_i, \b_i$ in terms of $\a_i$}
To prevent the ``effectivization" problem, we start with Eqs. (\ref{P}) and (\ref{fgh1})
to solve $\m, f_1$ and $f_0$ in terms of $\a_3$.
First, we introduce constants of motion $s_i,~ i=1,4$ which are defined
\be
P(\l) = f^2 -gh =\l^4 -s_1 \l^3 + s_2 \l^2 - s_3 \l +s_4.
\label{s1234}
\ee
By inserting $f,g$ and $h$ in Eq. (\ref{fgh1}) into Eq. (\ref{s1234}), we can obtain
\ben
s_1 &\equiv& \sum \l_i =-2 f_1 \a_3 +\a_+ \a_- ( \m + \m^* ), \nonumber \\
s_2 &\equiv& \sum_{i<j} \l_i \l_j =2 f_0 \a_3 +f_1 ^2 +\a_+ \a_- \m \m^*, \nonumber \\
s_3 &\equiv& \sum_{i<j<k} \l_i \l_j \l_k = -2 f_1 f_0,~~s_4 \equiv
\prod \l_i =f_0 ^2.
\label{eqs1234}
\een
Now solving for $f_0, f_1, \m$ and $\m ^*$ in Eq. (\ref{eqs1234}), we obtain
\be
f_1 = {s_3 \over
2 \sqrt s_4},
~f_0 = -\sqrt s_4,
~\m, \m ^* = {1 \over 2 (1 -\a_3 ^2 ) s_4 } \left( \sqrt s_4 s_3 \a_3
+s_1 s_4 \mp \sqrt{s_4 R} \right),
\label{f1f0}
\ee
where
\ben
R &=& (x+\l_2 \l_3 +\l_1 \l_4) (x+\l_1 \l_3 +\l_2 \l_4 )
(x+\l_1 \l_2 +\l_3 \l_4 ), \nonumber \\
x &=& 2 \sqrt s_4  \a_3.
\een
Note that $f_i, i=0,1$ are  constants, which are consistent with Eqs. (\ref{eq11}) and (\ref{eqN}).
To obtain this solution, we use Mathematica for symbolic manipulation. 
In addition, Mathematica was used to check
various formulae appeared in this paper. For example, it was used to check that Eqs. (\ref{f3}) and (\ref{f+}) 
satisfies the equation of motion
(\ref{vortex}).

The definition of $\g_i, i=1,3$ in Eq. (\ref{UV}) together with Eqs. (\ref{eq11}), (\ref{eqp}) and (\ref{eqs1234})
give
\ben
\g_3 &=& -i \a_+ \a_- ( \m + \m^* +2 f_1 /\a_3 ) \a_3 = -i (2 f_1 /\a_3 + s_1) \a_3 \equiv t \a_3, \nonumber \\
\g_+ &=& \{2i (\m+ \a_3 f_1) -i \a_+ \a_- (\m + \m^*) \} \a_+ = i(2 \m -s_1) \a_+ \equiv y \a_+ ,
\label{g3p}
\een
and $\g_- = -y^* \a_- $.
Note that these results are consistent with the third equations of Eqs. (\ref{eqNN}) and (\ref{eqN}).

Now, we require that $\a_3$ becomes a traveling wave in the 1-phase solution, i.e., a function of $w \equiv z + v \zb$, 
where $v$ is a constant, requiring $
\pb \a_3  = v \pp \a_3$. The definition of $\a_3, \b_3$ in Eq. (\ref{UV}) gives $\pb \a_3 = \pp \b_3$,
which in turn becomes
\be
\b_3 = v \a_3 +c(\zb).
\label{Vd}
\ee
To fix $c(\zb)$, we need following steps.
The fourth equations of Eqs. (\ref{eqNN}) and (\ref{eqN}) together with the results in Eqs. (\ref{g3p}), (\ref{Vd}) give
\be
\b_+ =\{f_1 ^2 /\a_3 ^2 +  \m f_1 /\a_3 - f_0 /\a_3 +v+c(\zb ) /\a_3  -{i \over 2} (f_1 /\a_3 +\m) t  -3 \t (f_1/ \a_3 + \m ) \} \a_+ 
\equiv u \a_+,
\label{bp}
\ee
and $\b_- = u^* \a_- $.
Now, the first equation of Eq. (\ref{eqN}) becomes
\be
\pb \a_3 ={i \over 2} \a_+ \a_- ( u \m^* - u^* \m).
\ee
Now, the relation $\pb \a_3 = v \pp \a_3$  with Eqs. (\ref{eq11}) and (\ref{bp}) gives
$c(\zb ) =c$ is a constant and
\be
c=f_0 + f_1 s_1 /2 +3 \t f_1.
\label{cff}
\ee
Finally, inserting all these results into the first equation of Eq. (\ref{eqNN}), we can obtain 
\be
\pb \m = -i (\m^2 + 3 \t \m +{i \over 2} \m y +u) (\a_3 \m^2 +f_1 \m +f_0) = v \pp \m,
\ee
which shows $\m$ is also a function of $w=z + v \zb$ only.

\subsection{Imposing the conditions $W=3 \t^2 -\k^2 /2 $ and $\t =$ constant}
To find the value of $v$, we need to impose the condition $W=3 \t^2 -\k^2 /2 $, i.e., $\Tr \pp r \pb r=6 \t^2-\k^2$.
First, we calculate the curvature $\k$. Using Eqs. (\ref{tn}), (\ref{UV}), (\ref{eq11}), (\ref{eqp}), (\ref{eqs1234}),  we find
\be
2 \k^2 = \Tr (\pp^2 r)^2 = -{1 \over 2} s_1 ^2 +2 s_2 -4 \a_3 f_0.
\label{curv}
\ee 
Similarly, we can obtain
\be
\Tr \pp r \pb r=2 v +2 \a_3 f_0- s_1 ^2/2,
\ee
which together with Eq. (\ref{curv}) gives
\be
v={3 \over 8} s_1 ^2 -{1 \over 2} s_2 +3 \t^2 +{3 \over 2} \t s_1.
\label{V}
\ee

Now, we calculate the torsion and impose the condition that the torsion is a constant.
We first find $\{ \bf{t, n, b} \}$ vectors as following,
\ben
t_i \s_i &=& \pp r = \pmatrix{\a_3 & \a_+ \cr \a_- & -\a_3}, \nonumber \\
n_i \s_i &=& {1 \over \k} \pp t_i ~\s_i={i \over \k} \pmatrix{  \a_- \a_+ (\m^* -\m)/2 & \a_+ (\a_3 \m +f_1) \cr
 -\a_- (\a_3 \m^* +f_1) & - \a_- \a_+ (\m^* -\m)/2},\nonumber \\
b_i \s_i &=& ({\bf n} \times {\bf t} )_i \s_i ={1 \over \k} \pmatrix{\a_3  s_1 /2 +f_1 & \a_+ (s_1 / 2 -\m) \cr
\a_- (s_1 / 2 -\m ^*) & -\a_3 s_1 /2 -f_1}.
\label{tnb}
\een
Inserting the result in Eq. (\ref{tnb}) into the equation $\pp b_3 = -\t n_3$ gives (use Eqs. (\ref{curv}) and (\ref{eq11}))
\be
({s_1 \over 2} +\t ) \k^2 +f_0 ({s_1 \over 2} \a_3 +f_1) =0,
\ee
which can be satisfied with a constant $\t$ by choosing  (use Eq. (\ref{curv}))
\be
s_1 = -4 \t,~~s_3 = 2 \t (4 \t^2 -s_2).
\label{condt}
\ee
We check that the curvature $\k$ given by Eq. (\ref{curv}) becomes the solution of the mKdV equation
when $s_1, s_3$ satisfies the conditions in Eq. (\ref{condt}).
Using Eq. (\ref{eqs1234}), the condition in Eq. (\ref{condt}) can be expressed in terms of $\l_{R1}, \l_{R2},
\l_{I1}, \l_{I2}$ where
$\l_j = \l_{Rj} +i \l_{Ij} , \l_{j+2} = \l_{Rj} -i \l_{Ij},~j=1,2$.

\vspace{1cm}
type-A : $\l_{R1}=\l_{R2}= -\t$, $\l_{I1}$ and $\l_{I2}$ are arbitrary,

type-B  : $\l_{R1}=-\l_{R2} -2 \t, ~\l_{I1}=\l_{I2}$.

\vspace{1cm}
Thus there exist two types of curves of the mKdV equation, which require different 
choice of the main spectra $\l_i$. Each type of main spectra has two independent parameters for curves of torsion $\t$.
They are $\l_{I1}$ and $\l_{I2}$ for the Type-A, while they are $\l_{R1}$ and $\l_{I1}$ for type-B.
In addition, we require the condition on $v$ in Eq. (\ref{V}).

\subsection{Derivation of $r_3$}
Using all these results, we can obtain the equation for $\a_3$,
\be
{d \a_3 \over dw} ={i \over 2 \sqrt s_4} \sqrt{R}.
\label{da3}
\ee
This equation can be integrated in terms of Weierstrass'  ${\cal P}(u, g_2 , g_3 )$ function.
As far as Weierstrass elliptic functions are involved, we employ terminology and
notation of \cite{jms} without further explanations. Explicitly
\be
\a_3 = {1 \over 2 \sqrt{s_4}} \left(-{s_2 \over 3} - 4{\cal P} (w+w_3 , g_2 ,g_3) \right),
\label{ab3}
\ee
where $w_3$ is an integration constant and 
\ben
g_2 &=& {1 \over 12} s_2 ^2 -{1 \over 4} s_1 s_3 + s_4, \nonumber \\
g_3 &=& {1 \over 216} s_2 ^3 -{1 \over 48} s_1 s_2 s_3
+{1 \over 16} s_3 ^2 -{1 \over 6} s_2 s_4 +{1 \over 16} s_1 ^2 s_4.
\een
The integration constant $w_3$ is determined by the initial condition, which we shall choose as follows;
${\cal P} (w_3 )=e_3$ at $w=0$, where $e_3$ is the smallest root of the equation $4 z^3 -g_2 z -g_3 =0$.
This condition guarantees that $|\a_3 | \le 1$.
Other two roots are denoted by $e_1$ and $e_2$ with $e_1 >e_2 >e_3$. 
$w_3$ as well as $w_1$ are called the half period  of the $\cal P$ function.  
They satisfy ${\cal P} (w_1 ) = e_1, {\cal P} (w_2 )=e_2, e_1 +e_2 +e_3 = w_1 +w_2 +w_3 =0$.
Especially,
\be
e_1 = {1 \over 4} (\l_1 \l_3 + \l_2 \l_4)-{s_2 \over 12},~
e_2 = {1 \over 4} (\l_1 \l_4 + \l_2 \l_3)-{s_2 \over 12},~
e_3 = {1 \over 4} (\l_1 \l_2 + \l_3 \l_4)-{s_2 \over 12}.
\label{e123}
\ee

Now using $\a_3 =\pp r_3$ and $\b_3 = \pb r_3$ in Eq. (\ref{Vd}), we can obtain (use Eq. (\ref{cff}))
\ben
r_3 &=& \int \a_3 dw+{s_3 s_1-4 s_4 +6 \t s_3 \over 4 \sqrt{s_4}}  \zb \nonumber \\
&=& {2 \over \sqrt{s_4} } \{ {\cal \z} (w+w_3 , g_2 , g_3) - {s_2 \over 12 } w
+{1 \over 8}(s_3 s_1- 4 s_4 +6 \t s_3) \zb -\h_3 \},
\label{f3}
\een
where ${\cal \z} (u, g_2 ,g_3 )$ is the Weierstrass' zeta function and the integration constant $\h_3$ is
taken to be ${\cal \z} (w_3, g_2, g_3 )$. (We will use the notation $\h_i ={\cal \z} (w_i, g_2, g_3 ), i=1,3$
in the following.)

\subsection{Derivation of $r_+$}
Now we try to obtain the solution $r_+ = r_1 -i r_2$ of the mKdV-type curve. The procedure
is similar as in \cite{shin1,shin2}, and we will only describe some important steps
of derivations. More details including the identities of Weierstrass' functions needed in 
the derivation can be found in \cite{shin1,shin2}.
Using Eqs. (\ref{g3p}), (\ref{Vd}), (\ref{bp}), we can express the second equation of Eq. (\ref{eqNN}) as
\be
\pb \a_+ = v \pp \a_+ + i(s_1 /2 +3 \t) f_0 \a_+ ,
\ee
which, with the help of Eq. (\ref{eqp}), gives
\be
\a_+ = \exp \{-i \sqrt{s_4} (s_1 /2 +3 \t) \zb \} ~\tilde \a_+ (w),
\label{abp1}
\ee
where $\tilde \a_+$ satisfies the following differential equation,
\be
{d \tilde \a_+ \over d w} = i \tilde \a_+ (f_1 +\a_3 \m ).
\label{abp}
\ee
Using Eqs. (\ref{f1f0}), (\ref{da3}), (\ref{ab3}), we can integrate Eq. (\ref{abp}) to obtain (see more details in \cite{shin1,shin2})
\be
\tilde \a_+ = 2{ i \over \sqrt s_4} {\s(w+w_3 +\k_1 ) \s(w+w_3 +\k_2 ) \over \s(\k_1 ) \s(\k_2 ) \s^2 (w+w_3 ) } 
\exp \{-\z(\k_1 ) w -\z(\k_2 )  w +\d \},
\label{apc}
\ee
where two constants $\k_1 , \k_2 $ are defined by the following relations 
(Please distinguish the constants $\k_1$ and $\k_2$ from the
curvature $\k$.),
\be
{\cal  P} (\k_1 ), {\cal P}(\k_2 ) = 
-s_2 /12 \pm  \sqrt{s_4}/2,
\label{pk1pk2}
\ee
\be
{\cal P} ' (\k_1 ), {\cal P} ' (\k_2 )  = - {\sqrt s_4 \over 2} {{d \a_3 \over dw}~\big|}_{\a_3 =\mp 1} 
={- {i \over 4} \sqrt R ~\big|}_{\a_3 =\mp 1}=-{i \over 4} (s_1 \sqrt s_4 \mp s_3 ).
\label{pk1pko}
\ee
The integration constant $\d$ can be fixed by requiring $\a_3 ^2 +|\a_+ |^2 =1$.
Explicit calculation of this requirement, especially at $w=0$, gives
$\d=-\h _3 (\k_1 +\k_2 )$.

We now derive $r_+$. Using $\pp r_+ =\a_+ $ in Eq. (\ref{UV}), we obtain
\be
r_+ = \int \a_+ dz +M(\zb)= \exp \{-i \sqrt{s_4} (s_1 /2 +3 \t) \zb \} \int \tilde \a_+ dw +M(\zb),
\label{Fbp}
\ee
where $M(\zb)$ is a function to be determined.
To evaluate the integration explicitly as well as to determine $M(\zb)$, we 
substitute Eq. (\ref{Fbp}) into the following relation,
\be
\pb r_+ = \b_+ = u \a_+ =(v-s_1 \m/2 -3 \t \m ) \a_+.
\ee 
It then gives $M(\zb)=0$ and (see more details in \cite{shin1,shin2})
\ben
r_+
&=& {2i \over s_4} \{\z(w+w_3 +\k_1 )-\z(w+w_3 +\k_2 )-\z(\k_1 )+\z(\k_2 ) \} \nonumber \\
&\times&  {\s(w+w_3 +\k_1 ) \s(w+w_3 + \k_2 ) \over \s(\k_1 ) \s(\k_2 )\s^2 (w+w_3 )}\nonumber \\
&\times& \exp \{-\z(\k_1 )w
-\z(\k_2 )w-\h_3 (\k_1 +\k_2 ) -i \sqrt{s_4} (s_1 /2 +3 \t) \zb \}.
\label{f+}
\een

Eqs. (\ref{f3}), (\ref{f+}) are the main results of the present paper. The space curves of the mKdV equation
are described by taking two types of the main spectra $\l_i, i=1,4$ shown in section 3.4. 
They are curves which have a constant torsion $\t$ and their curvature $\k$ is given by Eqs. (\ref{curv})
and (\ref{ab3}).

\section{Special Cases}
In this section, we study some special cases of the obtained solution
by taking specific values on the main spectra $\l_i$. 
The Weierstrass functions reduce to simple forms (for example, sinusoidal or elliptic functions) 
on these special limits.
These cases contain the straight line, the Kelvin wave and the 1-solitonic
curve, as well as
plane curves in terms of Jacobi functions, and the closed ring.

\subsection{The straight line}
The simplest solution that corresponds to the straight line is obtained by taking
$\l_{I1} =\l_{I2} =0$ in the type-A curve (or $\l_{R1} =-\t, \l_{I1} =0$ in the type-B curve), i.e.,
$\l_1 =\l_2 =\l_3 =\l_4 =-\t$ (real).
In this case, $g_2 =g_3 =e_3 =0$. 
The Weierstrass functions in this limit are given by
${\cal P} (u,0,0) =1/u^2, ~\z(u,0,0)=1/u, ~\s(u,0,0) =u$. And $w_3=\k_1 = \infty$.
Using these relations, we can easily obtain $r_3 =-z+3 \t^2 \zb, r_+ = r_1 -i r_2 =0$.
The rotational and translational symmetry of the vortex equation can give more general configuration
of the straight line \cite{shin1}.

\subsection{The Kelvin wave}

The next example, known as the Kelvin wave in fluid mechanics \cite{betchov}, corresponds to taking 
$\l_{I1} =0$ in the type-A curve, i.e.,
$\l_1 =\l_3 =-\t, \l_2 =-\t+i \g, \l_4 =-\t -i \g$.
In this case, $g_2 = { 1 \over 12} \g^4, g_3 ={1 \over 216} \g^6$. As $\D \equiv g_2 ^3  -27 \g_3 ^2 =0$,
the Weierstrass functions are given in simple forms,
\ben
{\cal P} (u) &=&-\g^2/12 +\g^2 \csc^2 (\g u/2)/4, \nonumber \\
\z(u)&=& \g^2 u/12
+ \g \cot ( \g u/2)/2, \nonumber \\
\s(u) &=&2 \exp ( \g^2 u^2 /24) \sin (\g u/2) / \g.
\een
And $e_2 =e_3 = -\g^2/12,
~w_3 =i \infty$,
\be
\sin (\g \k_1 /2), \sin (\g \k_2 /2) = {i \g \over \sqrt{ 2 \t^2 \pm 2 \t  \sqrt{\t^2 +\g^2 }}},
\ee 
\be
\cot (\g \k_1 /2), \cot (\g \k_2 /2) =-{i \over \g} \left(\sqrt{\t^2 +\g^2} \pm \t \right).
\ee 
Now, a straightforward calculation gives
\ben
r_3 &=& {1 \over \sqrt{\t^2 +\g^2}} \{\t z +3 (\t  \g^2 /2 +\t^3) \zb \},
\nonumber \\
r_+ &=& i{\g \over \t^2 +\g^2  } \exp \{ i \sqrt{\t^2 +\g^2 } (z  +\t^2 \zb -\g^2 \zb /2 )\}.
\een
This curve is a helix ($r_+ = r_1 -i r_2$). When $\t=0$, it describes a circular ring
having a constant radius and rotating on the $xy$-plane. The parameter $\g$ is the 
curvature $\k$ of the curve, i.e., $\k = \g$, while $\t$ is the torsion of the curve.
The three orthogonal vectors are given by
\ben
{\bf t} &=&{1 \over \sqrt{\t^2 +\k^2}  } (-\k \cos \T,
\k \sin \T, \t), \nonumber \\
{\bf n} &=& (\sin \T,
\cos \T, 0), \nonumber \\
{\bf b} &=&{1 \over \sqrt{\t^2 +\k^2}  } (\t \cos \T,
-\t \sin \T, \k),
\een
where $\T=\sqrt{\t^2 +\k^2 } (z  +\t^2 \zb -\k^2 \zb /2 )$.

\subsection{The Hasimoto 1-solitonic curve}
To obtain the curves of 1-soliton from our periodic solution, we take
$\l_{I1} =\l_{I2}$ in the type-A curve (or $\l_{R1} =-\t$ in the type-B curve), i.e.,
$\l_1 =\l_2 =-\t+i \g,~~ \l_3 =\l_4 =-\t-i \g$.
In this case $g_2 = { 4 \over 3} \g^4, g_3 =-{8 \over 27} \g^6$ and 
$w_3 =i \p/(2 \g), ~e_3={\cal P} (w_3)=-2 \g^2/3, ~\h_3 =\z (w_3) =-i \g \p/6,
~\s(w_3 )=i\exp(\p^2/24) /\g$.
The Weierstrass functions are given by
\ben
{\cal P} (w) &=&-2 \g^2/3 +\g^2 \coth^2 (\g w), \nonumber \\
\z(w)&=&-\g^2 w/3+\g \coth (\g w), \nonumber \\
\s(w) &=&\exp (- \g^2 w^2 /6) \sinh (\g w) /\g.
\een 
And $\k_1 =\infty, ~\sinh \g  \k_2= -i \g / \sqrt {\t^2 +\g^2 }, ~\coth \g  \k_2 =i \t /\g$.
It then gives
\ben
r_3 &=& -z -3 \t^2 \zb+{2 \g \over \t^2 +\g^2 } \tanh  ( \g z-\g^3 \zb), \nonumber \\
r_+ &=& -{2i \g \over \t^2 +\g^2} \sech ( \g z-\g^3 \zb) \exp (-i \t z -i\t^3 \zb).
\een

The curvature of the curve is 
$\k =2  \g \sech(\g z-\g^3 \zb)$. It satisfies the mKdV equation in Eq. (\ref{mkdv}).
The three orthogonal vectors can be easily constructed. For example,
\ben
{\bf t} &=& {2\g \sech \O \over \t^2 +\g^2  } (  \g \sin \T \tanh \O  
-\t \cos \T ,
 -   \g \cos \T \tanh \O -\t \sin \T , \nonumber \\
 && \g \sech \O -{ \t^2 +\g^2 \over 2 \g} \cosh \O ),
\een
where $\O=\g z-\g^3 \zb, \T=\t z +\t^3 \zb$.
In Figure 1, we show an example of the 1-solitonic curve  with parameters $\g=1/3$ and
(a) $\t=1$, (b) $\t=3$. 
\begin{figure}
\leftline{\epsfxsize 2.5 truein \epsfbox {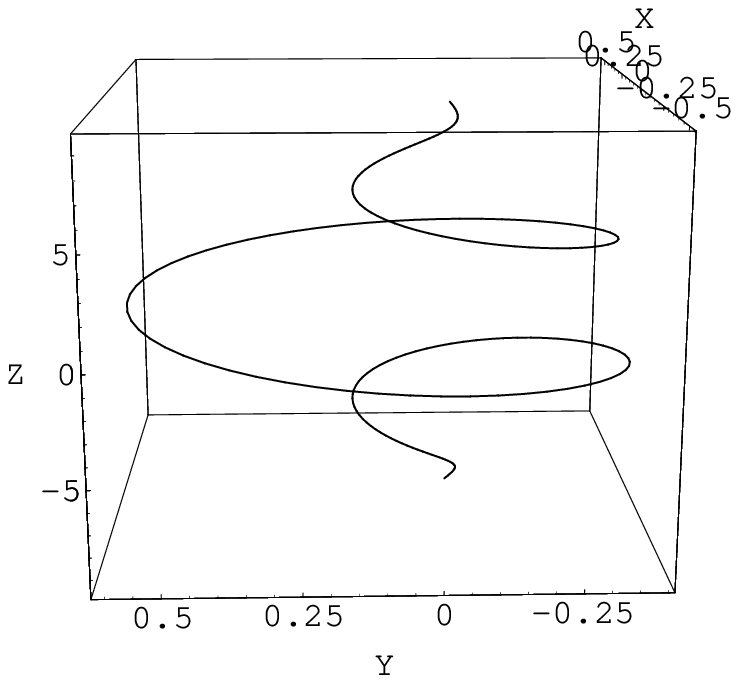}}
\vglue -2.15 in
\rightline{\epsfxsize 2.5 truein \epsfbox {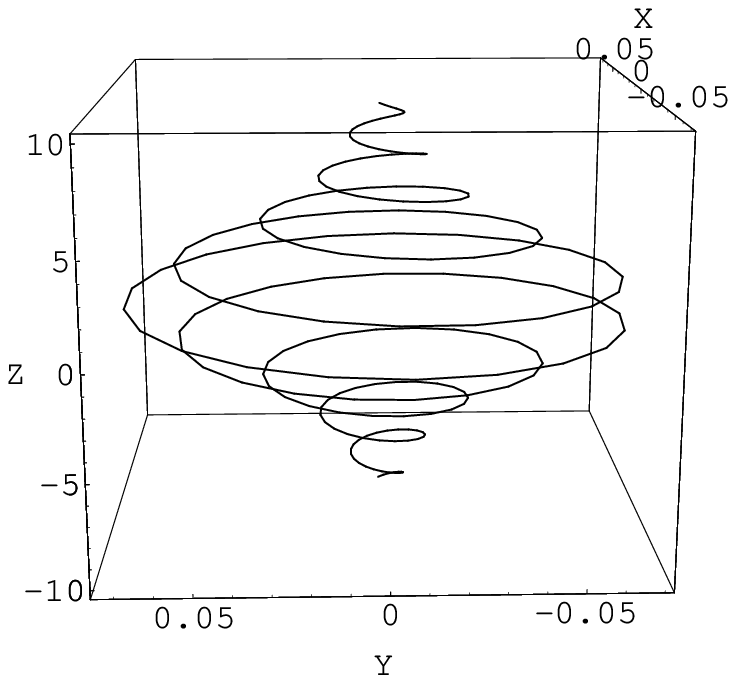}}
\vglue 0.5in
\caption{Typical 1-soliton curves drawn for $-10<z<10$, (a) with $\tau =1$, (b)
with $\t=3$. The parameter of the curves is $\g=1/3$, drawn at $\zb=0$. }
\end{figure}

\subsection{The rigid vortex filament}
When we take $\l_{I2}=\l_{R2}+\t$ in the type-B curve, i.e.,
$\l_1= \l_3 ^* =-a-2\t +i (a+\t) , \l_2 = \l_4 ^* =a + i (a+\t)$ for arbitrary $a$, then $v =0$ (use Eq. (\ref{V}) )
and $w=(z- v \zb) \rightarrow z$. 
Thus it describes a rigid filament of fixed shape, which rotates around the $z$-axis
with constant angular velocity (=$-\t \sqrt{4 (a +\t) ^4 + \t ^4}$) and moves
along the $z$-axis with constant velocity (=$-(4 (a +\t) ^4 + 3 \t ^4) /\sqrt{4 (a +\t) ^4 + \t ^4}$).
This configuration could be easily identified in experiments.

\subsection{Plane curves in terms of Jacobi functions}
\begin{figure}
\leftline{\epsfxsize 2.5 truein \epsfbox {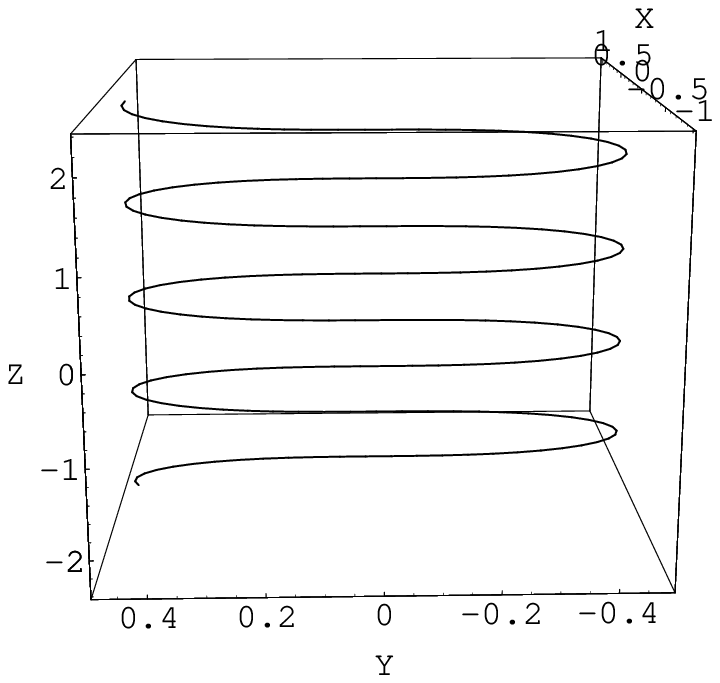}}
\vglue -2.15 in
\rightline{\epsfxsize 2.5 truein \epsfbox {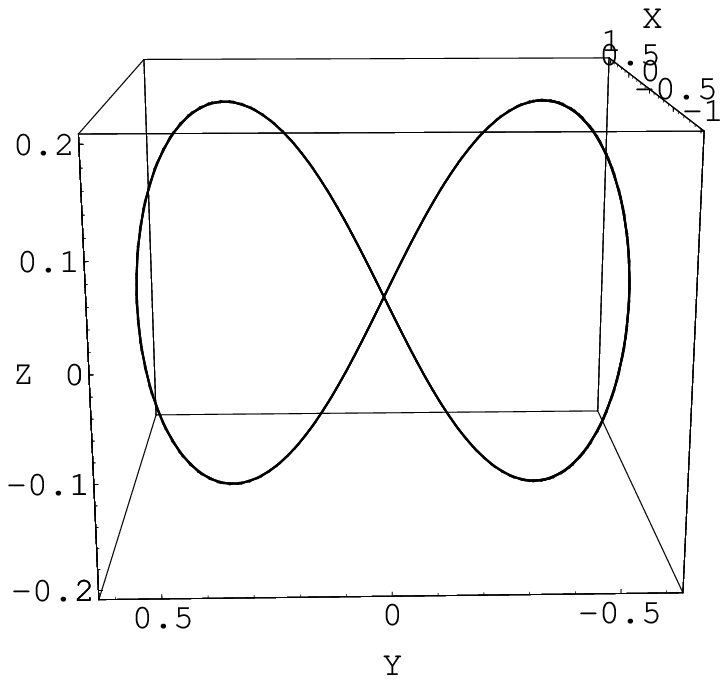}}
\vglue 0.5in
\caption{(a) Open plane curve
for $-10 < z <10$  and $\zb=0$. (b) Closed plane curve showing the famous figure-8 shape.}
\label{plane}
\end{figure}

Another reduced form of our solution is obtained by choosing $\k_1 =w_1, \k_2=w_3$.
Then ${\cal P} (\k_1)={\cal P} (\o_1)=e_1, {\cal P} (\k_2)={\cal P} (\o_3)=e_3$, and
Eqs. (\ref{pk1pk2}) and (\ref{e123}) fix the main spectra as
$\l_1 =\l_3 ^* =\r \exp(i \q), \l_2 =\l_4 ^* =-\r \exp(-i \q)$.
Note that it corresponds to taking $\t=0$ in the type-B curve, i.e., $\l_{R1} =\r \cos \q, \l_{I1} = \r \sin \q$ .
The Weierstrass $\s$ function
is reduced to the Jacobi's elliptic function under this limit as following,
\ben
{\s(w+w_3 +\k_1 ) \s(w+ w_3 +\k_2 ) \over \s ^2 (w+w_3 )}
=\exp\left( {\h_1 \over 2 w_1 } (w_1 ^2 +w_3 ^2 -2 w_2 w_3 -2 w_2 w) \right) \nonumber \\
\times q^{-3/4} \exp \left(-i {\p \over 2 w_1 } w \right) \sqrt{k \over k'} {\rm sn}(\sqrt{e_1 -e_3} w) 
{\rm dn}(\sqrt{e_1 -e_3} w).
\een
Using the relation $\h_1 w_2 =-i \p/2 +\h_2 w_1 $ (Legendre's relation) and
$k=\sqrt{e_2 -e_3 / e_1 -e_3 } = \sin \q, ~k' = \cos \q, q= \exp(i \p {w_3 / w_1})$,
we can obtain
\ben
r_+ &=& 2i \sin \q ~{\rm cn} (\r w)/\r, \nonumber \\
r_3 &=& \int \left( -1+2 {\rm dn}^2(\r w) \right) dw -\r^2 \zb = -w +2E(\r w|k)/\r -\r^2 \zb,
\een
where $E$ is the incomplete elliptic integral of the second kind. 
The curve lies on the $yz$-plane, and has the curvature $\k=2k\r {\rm cn} \{z-\r^2 (2 k^2-1) \zb\}$
and torsion $\t=0$. 
Figure \ref{plane} show examples of the plane curve. Fig. 2(a) shows an open plane curve
which is obtained by taking $\l_1 =\l_3 ^* = (3+3i)/{\sqrt 2}, \l_2 =\l_4 ^* = (-3+3 i)/{\sqrt 2}, \t=0$.
Fig. 2(b) shows the famous figure-8 shape (closed plane curve) in \cite{love,wadati}, which is obtained
by taking special value $\q =1.14$. The condition for closed curves will be discussed
in the following section.

\subsection{Closed curves}
The characteristics of closed curves can be obtained from the following quasi-periodicity properties
of Weierstrass' elliptic functions; 
\ben
\z(u+2 w_i) &=& \z(u)+2 \h_i, \nonumber \\
\s(u+2 w_i) &=& -\exp\{2 \h_i (u+w_i) \}\s(u).
\label{pseudoperiodicity}
\een
In our problem,
$w_1$ is real, while $w_3$ is pure imaginary.
Thus the physical characteristics of closed curves are described by the period $\D w=2 w_1$. 
After this period $2 w_1$, $r_+$ obtains an additional factor 
\be
\exp \{ 2 \h_1 (\k_1 +\k_2 ) -2 \z (\k_1 ) w_1  -2 \z (\k_2 ) w_1 \}.
\label{factor}
\ee
Thus a necessary condition for a closed curve
, i.e., $r_+ (z=2 m w_1 ) =r_+ (z=0)$ , is
\be
\h_1 (\k_1 +\k_2 ) - w_1 \{\z(\k_1 ) +\z (\k_2 ) \} =i {n \over m} \p,
\label{cond1}
\ee
where $m, n$ are arbitrary integers. Another condition comes from the closedness of the $z$-coordinate, i.e.,
$r_3(z=2 m w_1 ) = r_3 (z=0)$, which is
\be
-{s_2 \over 12 \sqrt s_4 } w_1 +{1 \over \sqrt{s_4} } \h_1=0,
\label{cond2}
\ee
i.e., $s_2 w_1 = 12 \h_1$.

The quasi-periodic property of the closed curves along the time is following.  During a time period
$\D \zb =2 w_1 /v$, the ring returns to its original shape, but rotates around the $z$-axis
by an angle
\be
2 {n \over m} \p -\sqrt{s_4} (s_1 +6 \t) {w_1 \over v},
\ee
and moves a distance
\be
{s_3 s_1 -4 s_4 +6 \t s_3 \over 2 \sqrt{s_4}} {w_1 \over v}
\ee
along the $z$-axis.

\begin{figure}
\leftline{\epsfxsize 1.8 truein \epsfbox {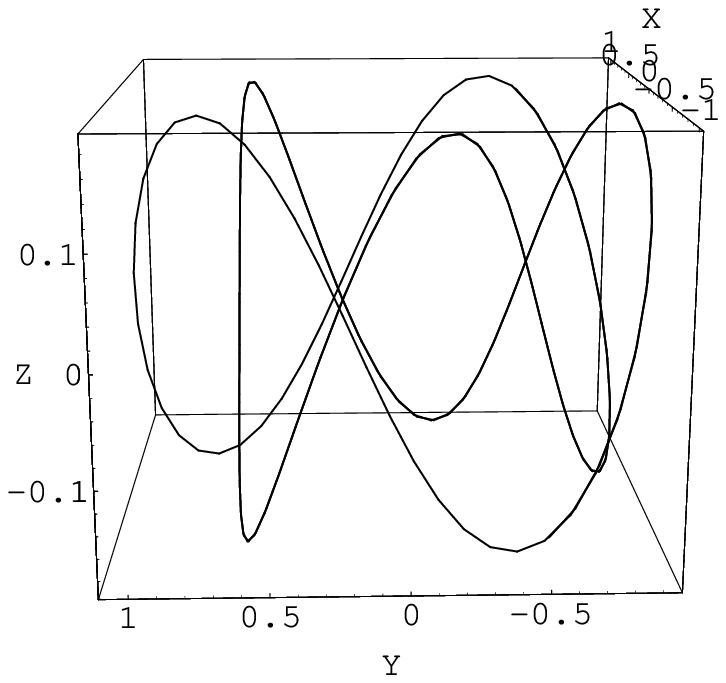}}
\vglue -1.6 in
\centerline{\epsfxsize 1.8 truein \epsfbox {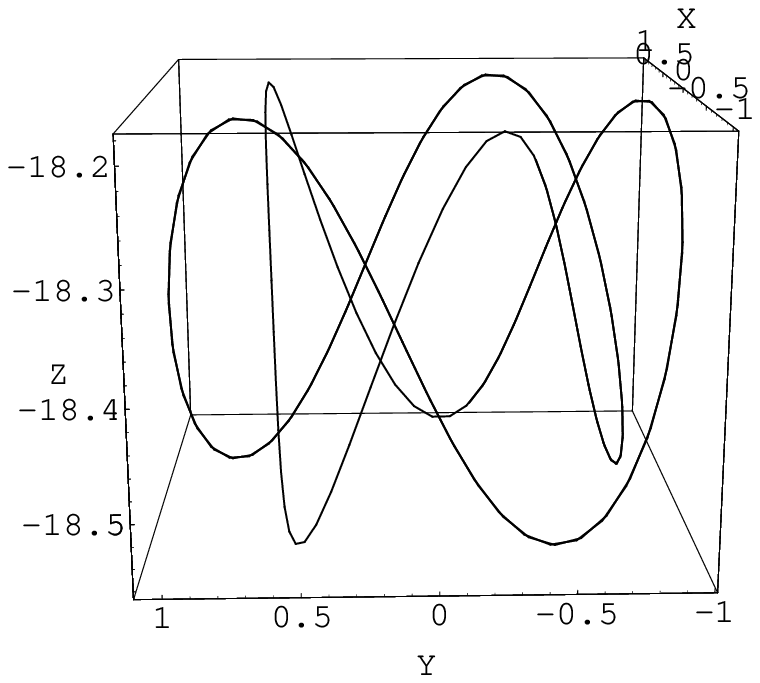}}
\vglue -1.6 in
\rightline{\epsfxsize 1.8 truein \epsfbox {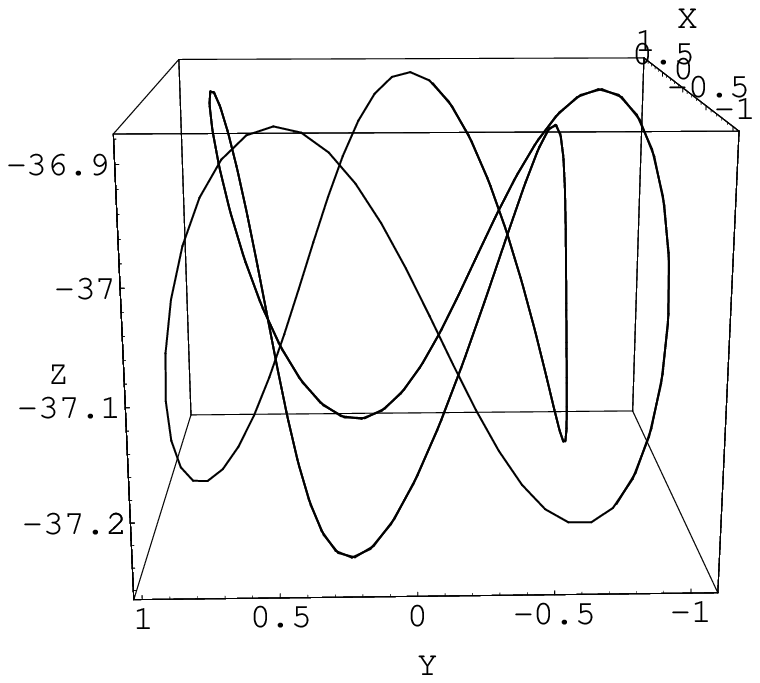}}
\vglue 0.2in
\caption{A typical closed ring plotted for $-10<z<10$ with parameters
$\l_1 =\l_3 ^* =-2.0653+1.171 i, \l_2 =\l_4 ^* = 0.0653+1.171 i, \tau =1$ (a) at $\zb=0$, 
(b) at $\zb=5$,  (c) at $\zb=10.09$ (quasi-period time).}
\label{figclo}
\end{figure}

The two conditions, Eqs. (\ref{cond1}) and (\ref{cond2}), can be solved numerically using a software package 
like Mathematica. One possible solution is $\l_1 =\l_3 ^* =-2.0653+1.171i, \l_2 =\l_4 ^* =0.0653+1.171i, \tau =1$,
which belongs to type-B curves. It satisfies the conditions (\ref{cond1}) and (\ref{cond2}) with
$n=2, m=5$ ($w_1 =-1.198, v=-0.2374$).  Figure \ref{figclo} shows the motion of a closed curve at 
(a)  $\zb=0$, (b) $\zb=5$, (c) $\zb=10.09$ (=$\D \zb$, quasi-period time), respectively.
During a time period $\D \zb$, the ring rotates -25.6 radian  and advances -37.07 along the $z$-axis.
The $z$ period is $10 w_1=-11.98$.
We can't find the type-A  curve that satisfies the closedness conditions in Eqs. (\ref{cond1}) and (\ref{cond2}).

In the case of figure-8 shape (closed plane curve), the condition  $\k_1 =w_1, \k_2=w_3$
satisfies Eq.  (\ref{cond1}) with $n=1, m=2$ (Legendre's relation.). 
The second condition (\ref{cond2}) becomes
$6 \z(w_1) = -\r^2 w_1$, which is satisfied by choosing $\l_1 =\l_3 ^* =\r(1+ 2.177966 i),
\l_2 =\l_4 ^* =-\r(1- 2.177966 i)$, or $\q =1.14$ in the notation of section 4.5. Fig. 2(b) shows the
figure-8 shape curve.

\subsection{General vortex filament}
\begin{figure}
\leftline{\epsfxsize 2.5 truein \epsfbox {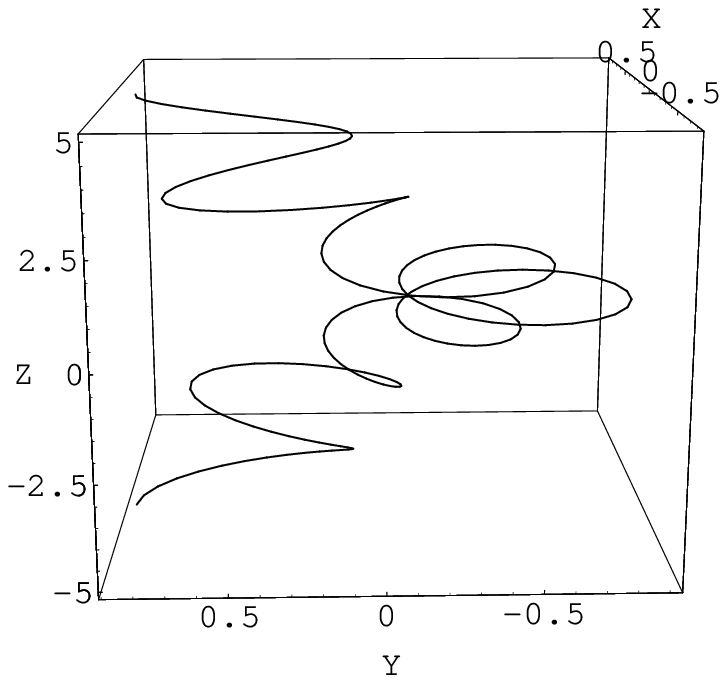}}
\vglue -2.15 in
\rightline{\epsfxsize 2.5 truein \epsfbox {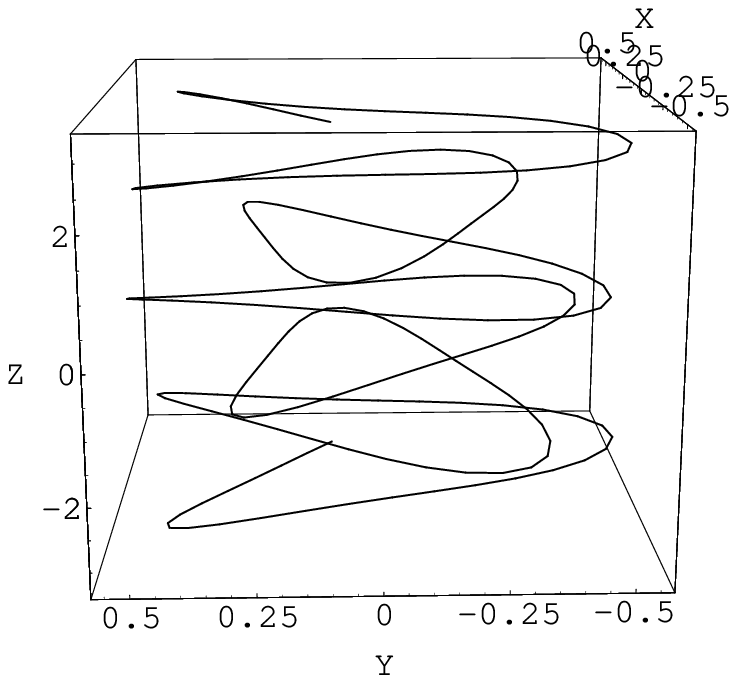}}
\caption{Typical open curves having $\tau =1$, drawn for $-10<z<10, \zb=0$, 
(a) type-A curve with $\l_1 =\l_3 ^* =-1+i, \l_2 =\l_4 ^* =-1+2i$,
and (b) type-B curve with $\l_1 =\l_3 ^* =-3+2i, \l_2 =\l_4 ^* =1+2i$. }
\label{gen}
\end{figure}

Most generally, the solution in Eqs. (\ref{f3}) and (\ref{f+}) describes an  open space curve 
with a constant torsion $\t$.
Figure \ref{gen}(a) shows an example (type-A curve), which we obtain using $\l_1 =\l_3 ^* =-1+i,
\l_2 =\l_4 ^* =-1+2 i, \t=1$. 
Figure \ref{gen}(b) shows another example (type-B curve), which we obtain using $\l_1 =\l_3 ^* =-3+2 i,
\l_2 =\l_4 ^* =1+2 i, \t=1$. These are examples of the most general configurations described
by the 1-phase quasi-periodic solution.

\section{Discussions}
In this paper, we present the calculation of the one-phase quasi-periodic solution of 
space curves described by the mKdV equation. The solution is explicitly given in terms 
of Weierstrass' elliptic functions. The main difference of the present calculation compared 
to that of previous works in \cite{kam2,shin1,shin2} is the appearance of constraints on 
the main spectra resulting from imposing two conditions. It then results in two types of 
curves, type-A and type-B curves. The first condition, $W=3 \t^2 -\k^2 /2$ in Section 2.4, is 
related with the internal parametrization of the curve. It does not determine the extrinsic 
form of the curve \cite{wadati}. This is consistent with the fact that the curve equation in 
Equation (\ref{vortex}) only fixes the form of $U$ and $V$ and does not fix $W$ in 
Equation. (\ref{dotr}). As is shown in Section 3.4, the condition on $W$ fixes the velocity 
$v$ of the motion of rigid curve. The second condition, $\t$ = constant, is also 
unique in the present calculation, and is not required in the previous calculations in 
\cite{kam2,shin1,shin2}. It was checked that the constraint on the main spectra is required 
for that $\k$ in Equation (\ref{curv}) satisfies the mKdV equation in Equation (\ref{mkdv}).

We study various configurations resulting from the degenerate limit of the quasi-periodic 
solution. They contain the Kelvin wave belonging to the type-A curve, the rigid vortex filament, 
the Jacobian plane curve, and closed curves belong to the type-B curve, and the straight 
line and the one-solitonic curve belonging both to the type-A and type-B curve simultaneously.

The simple case of zero-torsion was studied in various contexts such as the elastica problem 
and the computer vision problem. In this case, the fact that the curve lies on a plane was 
the main keystone in the derivation of the curve configurations. In the case of non-zero 
torsion, the convenient formalism of differential geometry seems difficult to realize except 
in some especially simple cases. The present derivation  heavily depends on the integrability 
structure of the mKdV equation. The equations for the one-phase solution in Section 3.2 are  
by no means simpler that the equations for the stationary curve configuration given in 
\cite{fuku} (Equations (3.32), (3.36), (3.38), and (3.39)). But there exists a standard procedure 
in solving these equations, shown in Section 3.2. Moreover, this procedure offers a 
generalization which can be used in calculating the $N$-phase periodic solutions.

Our formalism might be adapted to the curve problem of the non-zero and non-constant 
torsion problem. It is known that the spectral parameter of the integrable equation can be 
extended to be $z, \zb$-dependent without destroying the integrability \cite{burt,cal}. 
Thus it could result in space curves of non-constant torsion.
Explicit construction of curves with non-constant torsion is rare and difficult. Thus it remains 
as an interesting work for the future.

\vskip1cm

\noindent{\bf Acknowledgements}

This work was supported by Korea Research Foundation Grant (KRF-2003-070-C00011).

\end{document}